\documentclass[10pt]{article}
\usepackage{hyperref}
\usepackage{amsmath}
\usepackage[]{authblk}

\begin{document}

		\title{Non-relativistic Gravity in Entropic Quantum Dynamics\thanks{Presented at MaxEnt 2010, the 30th International Workshop on Bayesian Inference and Maximum Entropy Methods in Science and Engineering (July 4-9, 2010, Chamonix, France).
	}}

	\renewcommand\Affilfont{\small}

	\author{David T. Johnson\thanks{Email: \href{mailto:dave@dtjohnson.net}{dave@dtjohnson.net}}
	}
	\author{Ariel Caticha\thanks{Email: \href{mailto:ariel@albany.edu}{ariel@albany.edu}}
	}
	\affil{Department of Physics, University at Albany-SUNY, \\ Albany, NY 12222, USA}
	
	\date{}
	\maketitle
	\begin{abstract}
		Symmetries and transformations are explored in the framework of entropic quantum dynamics. Two conditions arise that are required for any transformation to qualify as a symmetry. The heart of this work lies in the application of these conditions to the extended Galilean transformation, which admits features of both special and general relativity. The effective gravitational potential that arises in a non-inertial frame through the strong equivalence principle arises naturally through an equivalence of information.
	\end{abstract}

	\section{Introduction}
		Science, and physics in particular, is concerned with ordering our experience. For any success in this endeavor, nature must adhere to some logical ordering. We refer to this logical scheme as the laws of nature. In turn, we collect observations in the form of measurements and find the relationship between these observations. These relationships take a mathematical form and are dubbed the laws of physics. A common perspective is that these laws and the laws of nature are one and the same; the mental picture painted by these physical laws is that of reality itself.
		
		In actuality, nature may be far more complicated. It may be that the laws of nature cannot be fully interpreted with the limited lens of our  comprehension \cite{stapp:1972}. The laws of physics would then describe reality in an indirect and incomplete manner and become rules for processing the limited information we have about reality. If this is indeed the case, then the methods of inference for limited information become very useful and very constraining.
		
		To this end, non-relativistic quantum mechanics has been formulated in informational terms \cite{caticha:2009,caticha:2010} as entropic quantum dynamics (EQD). This approach assumes that a particle is associated with hidden variables specified by a spatially-dependent probability distribution. Dynamics is introduced by allowing the particle to move in short but uncertain steps. Applying the principle of maximum entropy and introducing a suitable model of time results in a diffusion theory. Adding the final step of asserting energy conservation leads to the Schr\"odinger equation. The details of this procedure are reviewed in section \ref{sec:eqd}. Readers familiar with EQD, particularly \cite{caticha:2010}, need not dwell on this section.
		
		Once the framework of entropic dynamics is in place, we explore the very powerful concepts of symmetries and transformations. These concepts are central to physics and their role in EQD is no less significant. An observer or frame of reference is characterized by the specific information available to that observer. A symmetry under a transformation relating two observers means that, despite possessing different states of knowledge, the two observers describe the physical situation in equivalent ways. The concept of symmetry, and two conditions for that symmetry, are discussed in section \ref{sec:sieqd}.
		
		Knowledge that nature exhibits a particular symmetry comprises highly relevant information. Given two observers connected by a transformation, we wish to know how the various expressions in the two frames must be related for the transformation to qualify as a symmetry. In particular, we examine this question for the extended Galilean transformation (EGT) in section \ref{sec:tegt}. This generalization of the standard Galilean transformation to an arbitrary, accelerating frame is particularly interesting as it yields residues of special and general relativity \cite{greenberger:2001}. In fact, the effective gravitational potential that arises in a non-inertial frame emerges naturally.

	\section{Entropic Quantum Dynamics}
		\label{sec:eqd}
		Entropic quantum dynamics is built on the assumption that the motion of a particle is influenced by some unknown, unspecified hidden variables. Consider a particle at a position $x$ in a flat, three-dimensional configuration space $\mathcal{X}$ with metric $\gamma_{ab} = \delta_{ab}/\sigma^2$. (The scale factor $\sigma^2$ is included to allow generalization to multiple particles.) The hidden variables $y$ live in a space $\mathcal{Y}$ and are described by an unknown probability distribution $p(y|x)$.

		Additionally, we assume that small changes from one state to another happen and that large changes are the accumulation of many small changes. Consider a particle at an initial position $x$ moving to some unknown position $x'$. We apply the method of maximum entropy subject to the constraint that steps from one position to another must be small. The transition probability that maximizes the entropy is
		\begin{equation}
			\label{eq:raw_prob}
			P(x'|x) = \frac{1}{\zeta(x, \alpha)} \exp \left[ S(x') - \tfrac{1}{2}\alpha(x) \Delta\ell^2(x',x) \right] \ ,
		\end{equation}
		where $\zeta(x, \alpha)$ is the normalization \cite{caticha:2010}. $S(x)$ is the entropy of the hidden variables relative to an underlying measure $q(y)$ of the space $\mathcal{Y}$,
		\begin{equation}
			\label{eq:entropy}
			S(x) = -\int\! dy \ p(y|x) \log \frac{p(y|x)}{q(y)} \ .
		\end{equation}
		The particle moves in the direction of increasing entropy, with the Lagrange multiplier $\alpha(x)$ controlling the step size.
		
		\bigskip
		
		Time is introduced as a means of keeping track of the accumulation of changes. Individual steps are described by the transition probability $P(x'|x)$, but larger changes are the accumulation of many short steps. We therefore denote the probability of a particle's position at a given time $t$ as $\rho(x,t)$ and the probability at a later time $t' = t + \Delta t$ as $\rho(x',t')$. To properly model the Newtonian time of non-relativistic quantum mechanics, the interval of time $\Delta t$ must be constant. This goal is achieved if $\alpha(x,t)$ is chosen to be constant, $\alpha(x,t) = \tau/\Delta t$, where $\tau$ is a constant that ensures $\Delta t$ has units of time \cite{caticha:time}.
		
		Expanding the transition probability (\ref{eq:raw_prob}) about the maximum displacement $\Delta\bar{x}$ and substituting this definition for $\alpha(x,t)$ simplifies the description of motion. Letting $\Delta x = x' - x$, the transition probability becomes
		\begin{equation}
			\label{eq:approx_trans_prob}
			P(x'|x) \approx \frac{1}{Z} \exp \left[ -\frac{\tau}{2\sigma^2\Delta t} \delta_{ab} (\Delta x^a - \Delta \bar{x}^a)(\Delta x^b - \Delta \bar{x}^b) \right] \ ,
		\end{equation}
		where a displacement $\Delta x^a = b^a(x)\Delta t + \Delta w^a$. The drift velocity $b^a(x)$ is given by
		\begin{equation}
			\label{eq:future_drift}
			\langle \Delta x^a \rangle = \Delta\bar{x}^a = b^a(x)\Delta t \quad \text{where} \quad b^a(x) = \frac{\sigma^2}{\tau} \partial^a S(x) \ .
		\end{equation}
		The fluctuations $\Delta w^a$ are of order $\smash{\sqrt{\Delta t}}$ and dominate as $\Delta t \rightarrow 0$. As in Brownian motion, the trajectory is continuous but non-differentiable.
		
		\bigskip

		From diffusion theory, the accumulation of the small changes according to (\ref{eq:approx_trans_prob}) and (\ref{eq:future_drift}) are described by a Fokker-Planck (FP) equation,
		\begin{equation}
			\label{eq:fp}
			\partial_t \rho = -\partial_a(v^a\rho) \ ,
		\end{equation}
		provided the \emph{current velocity} is
		\begin{equation}
			v^a = b^a + u^a \ , \qquad \text{where} \qquad u^a \overset{\underset{\mathrm{def}}{}}{=} - \frac{\sigma^2}{2\tau} \partial^a \log \rho
		\end{equation}
		is the \emph{osmotic velocity}.
		The mean drift $b^a$ drives the probability flow up the entropy gradient, while the osmotic velocity $u^a$ drags it down the concentration gradient. Further, we can write the current velocity $v^a$ as a gradient as well,
		\begin{equation}
			\label{eq:untrans_current_velocity}
			v^a = \frac{\sigma^2}{\tau} \partial^a \phi \ , \qquad \text{where} \qquad \phi(x,t) = S(x) - \log \rho^{1/2}(x,t) \ .
		\end{equation}

		The dynamics described thus far is diffusion, not quantum mechanics. To construct a wave function $\smash{\Psi = \rho^{1/2}e^{i\phi}}$, we need to promote the phase $\phi$ to an independent degree of freedom. That is, we must allow the entropy field $S(x)$ to change in response to the dynamics. To specify the way in which $S(x,t)$ changes, we introduce an energy functional,
		\begin{equation}
			\label{eq:energy1}
			E[\rho,v] = \int \! d^3x \ \rho(x,t) \left( \tfrac{1}{2}mv^2 + \tfrac{1}{2}m u^2 + V(x,t) \right) \ ,
		\end{equation}
		where $m$ is called the \emph{mass}. (In general, one could use a different \emph{osmotic mass} $\mu$ for the osmotic velocity term. However, one may always tune units in such a way that $\mu = m$, thereby simplifying the dynamics \cite{caticha:2009}.) The field $V(x,t)$ represents an external potential. We also impose an energy conservation condition such that the energy changes at the rate
		\begin{equation}
			\label{eq:energy_cons}
			\dot{E} = \int \! dx \ \rho \dot{V} \ .
		\end{equation}
		When the potential is time-independent (i.e. $\dot{V} = 0$), the energy conservation condition is simply $\dot{E} = 0$.
		
		By requiring that the conservation condition hold for arbitrary choices of $\rho$ and $\phi$ and applying manipulations involving integration by parts and the FP equation, we derive a dynamical equation,
		\begin{equation}
			\label{eq:pde_phi}
			\eta \dot{\phi} + \frac{\eta^2}{2m}(\partial_a\phi)^2 + V - \frac{\eta^2}{2m} \frac{\nabla^2 \rho^{1/2}}{\rho^{1/2}} = 0 \ ,
		\end{equation}
		where the new constant $\eta$ satisfies $m = \tau\eta/\sigma^2$.

		Finally, we can combine the two coupled differential equations (\ref{eq:fp}) and (\ref{eq:pde_phi}) into a complex function $\smash{\Psi = \rho^{1/2} e^{i\phi}}$. Identifying $\eta = \hbar$ reproduces the Schr\"odinger equation,
		\begin{equation}
			i\hbar \frac{\partial\Psi}{\partial t} = -\frac{\hbar^2}{2m}\nabla^2\Psi + V\Psi \ .
		\end{equation}

	\section{Symmetry in Entropic Dynamics}
		\label{sec:sieqd}
		A symmetry is the inability to distinguish physical situations \cite{wigner:1967}. There is a symmetry between two physical situations when the laws of physics and the observations that are correlated by those laws are invariant -- at least in some limited way.
		From the viewpoint of entropic dynamics, a symmetry arises when two observers have \emph{equivalent} states of knowledge or information. In general, the information available to each observer is different; but despite this difference, they would describe the dynamics in an equivalent way.
		
		One key piece of information in EQD is the relation of hidden variables to the particle's position represented by the probability distribution $p(y|x)$. Different observers with different states of knowledge would not, in general, describe this relationship in the same way. That is, the probability distributions for the hidden variables in the two frames of reference are not necessarily equal. We wish to know what is the particular way in which the probabilities differ such that the transformation qualifies as a symmetry.

		Two conditions in EQD should be singled out as important requirements of symmetry. First, if two observers differ by a symmetry transformation, they must agree on the probabilities assigned for the transition between two corresponding points. That is, if an observer in $\mathcal{X}$ assigns the probability density $P(x'|x)$ to a transition from $x$ to $x'$, and another observer in $\smash{\widetilde{\mathcal{X}}}$ assigns $\tilde{P}(\tilde{x}'|\tilde{x})$ for the corresponding transition from $\tilde{x}$ to $\tilde{x}'$, then
		\begin{equation}
			\label{eq:sym_con1}
			\tilde{P}(\tilde{x}'|\tilde{x}) \, d^3\tilde{x}' = P(x'|x)\, d^3x' \ .
		\end{equation}
		This condition ensures equivalence of the dynamics in the two frames.
		
		The second condition requires that the predictions in the two frames of reference coincide. For any given time $t$ and corresponding $\tilde{t}$, the probabilities assigned to a particular position should be identical. That is,
		\begin{equation}
			\label{eq:sym_con2}
			\tilde{\rho}(\tilde{x},\tilde{t})\, d^3\tilde{x} = \rho(x,t)\, d^3x \ .
		\end{equation}
		These two symmetry conditions greatly constrain transformations.

	\section{The Extended Galilean Transformation}
		\label{sec:tegt}
		Invariance with respect to the Galilean transformation is a geometrical symmetry between frames which differ by a constant velocity. One can generalize this transformation to an arbitrary, translational acceleration known as the extended Galilean transformation \cite{rosen:1972,greenberger:1979}. This extended symmetry is of particular interest because it retains residual features of both special and general relativity in non-relativistic quantum mechanics.
		
		Consider a new observer describing the dynamics of the same particle in section \ref{sec:eqd}. In the new observer's frame of reference, the particle lives in a 3-dimensional space $\smash{\widetilde{\mathcal{X}}}$. We are not assuming a definition of time yet, only that the particle will take small steps. Accordingly, the transformation connecting the two spaces is $\tilde{x}^a = x^a + \xi^a(\theta)$, where $\xi$ is an arbitrary displacement and $\theta$ is a parameter that varies as the particle takes steps. (Inclusion of a static rotation is straightforward.) Both the metric and the volume element $d^3x$ are invariant as $\xi$ is spatially uniform.

		The introduction of dynamics in the transformed frame follows a very close parallel with the original derivation. Applying the method of maximum entropy subject to normalization and the constraint of small steps leads to a transformed transition probability $\smash{\widetilde{P}}(\tilde{x}'|\tilde{x})$ of the same form as (\ref{eq:raw_prob}) but with a new Lagrange multiplier $\tilde{\alpha}(\tilde{x})$ and a transformed entropy $\smash{\widetilde{S}}(\tilde{x})$. The transformed entropy represents the new observer's state of knowledge about the hidden variables.

		We wish to model the very same Newtonian time in $\smash{\widetilde{\mathcal{X}}}$ so that time flows not only at the same rate everywhere in space but at the same rate in every frame. Hence we define $\alpha(x,t) = \tilde{\alpha}(\tilde{x},\tilde{t}) = \tau/\Delta{\tilde{t}} = $ constant. Now we can identify the parameter $\theta$ with the time $\tilde{t}$ and specify the full transformation
		\begin{equation}
			\label{eq:trans}
			\tilde{x}^a = x^a + \xi^a(t) \ , \qquad \tilde{t} = t \ ,
		\end{equation}
		with derivatives,
		\begin{equation}
			\tilde{\partial}_t = \partial_t - \dot{\xi}^a \partial_a \ \qquad \text{and} \qquad \tilde{\partial}_a = \partial_a \ ,
		\end{equation}
		where $\tilde{\partial}_a = \partial/\partial \tilde{x}^a$ and $\dot{\xi}^a = \partial_t \xi^a(t)$.

		Upon substituting the definition of $\tilde{\alpha}$ into the transition probability and expanding about the maximum displacement $\Delta\tilde{\bar{x}}$,
		\begin{equation}
			\label{eq:transformed_approx_trans_prob}
			\widetilde{P}(\tilde{x}'|\tilde{x}) \approx \frac{1}{\widetilde{Z}} \exp \left[ -\frac{\tau}{2\sigma^2 \Delta \tilde{t}} \delta_{ab} (\Delta \tilde{x}^a - \Delta \tilde{\bar{x}}^a)(\Delta \tilde{x}^b - \Delta \tilde{\bar{x}}^b) \right] \ ,
		\end{equation}
		where the displacement $\Delta \tilde{x}^a = \tilde{b}^a(\tilde{x})\Delta \tilde{t} + \Delta \tilde{w}^a$. The drift velocity $\tilde{b}^a(\tilde{x})$ is given by
		\begin{equation}
			\label{eq:trans_future_drift}
			\langle \Delta \tilde{x}^a \rangle = \Delta\tilde{\bar{x}} = \tilde{b}^a(\tilde{x})\Delta \tilde{t} \quad \text{where} \quad \tilde{b}^a(\tilde{x}) = \frac{\sigma^2}{\tau} \tilde{\partial}^a \widetilde{S}(\tilde{x}) \ ,
		\end{equation}
		while the expectated fluctuations are unchanged by the transformation.

		\bigskip
		
		We require that the dynamics obey the symmetry conditions (\ref{eq:sym_con1}, \ref{eq:sym_con2}). The first condition asserts that the transition probabilities are invariant. Comparing these probabilities in both frames (\ref{eq:approx_trans_prob}, \ref{eq:transformed_approx_trans_prob}) implies that their exponents must be equal, as the normalization function $\smash{\widetilde{Z}} = Z$. This implies
		\begin{equation}
			\label{eq:sym1}
			\Delta \tilde{x}^a - \Delta \tilde{\bar{x}}^a = \pm (\Delta x^a - \Delta \bar{x}^a) \ .
		\end{equation}
		The negative sign is rejected as being inconsistent with the limit of $\xi \rightarrow 0$ where $\tilde{x} \rightarrow x$. The transformed displacement can be expressed in terms of the original displacement, $\Delta \tilde{x}^a = \Delta x^a + \Delta \xi^a$. By substituting this and the expressions for the mean displacements in both frames (\ref{eq:future_drift}, \ref{eq:trans_future_drift}) and rearranging, we have
		\begin{equation}
			\tilde{\partial}_a \left(\widetilde{S}(\tilde{x}) - S(x)\right) = \frac{\tau}{\sigma^2}\frac{\Delta \xi}{\Delta t} \ .
		\end{equation}
		Solving the differential equation for the entropy shift and taking the limit $\Delta t \rightarrow 0$ gives
		\begin{equation}
			\label{eq:entropy_shift_mostly}
			\widetilde{S}(\tilde{x}) - S(x) = \frac{\tau}{\sigma^2} \left( \dot{\xi}^a \tilde{x}_a +c(t) \right) \ ,
		\end{equation}
		where $c(t)$ is a constant of integration.

		The transformed drift and osmotic velocities relative to the original frame are
		\begin{equation}
			\label{eq:trans_velocities}
			\tilde{b}^a = b^a + \dot{\xi}^a \qquad \text{and} \qquad \tilde{u}^a \overset{\underset{\mathrm{def}}{}}{=} - \frac{\sigma^2}{2\tau} \tilde{\partial}^a \log \tilde{\rho} = u^a \ ,
		\end{equation}
		where the drift velocity follows from (\ref{eq:trans_future_drift}) and (\ref{eq:entropy_shift_mostly}). The invariance of the osmotic velocity follows directly the second symmetry condition (\ref{eq:sym_con1}). If we let $\tilde{v}^a = \tilde{b}^a + \tilde{u}^a$ be the transformed current velocity, the transformed FP equation is covariant,
		\begin{equation}
			\label{eq:trans_fp}
			\tilde{\partial}_t \tilde{\rho} = -\tilde{\partial}_a(\tilde{v}^a\tilde{\rho}) \ .
		\end{equation}
		Additionally, it follows from (\ref{eq:trans_velocities}) that $\tilde{v}^a = v^a + \dot{\xi}^a$. Finally, we can express the current velocity as a gradient,
		\begin{equation}
			\tilde{v}^a = \frac{\sigma^2}{\tau} \tilde{\partial}^a \tilde{\phi} \ , \qquad \text{where} \qquad \tilde{\phi}(\tilde{x},\tilde{t}) = \phi(x,t) + \frac{\tau}{\sigma^2} \left( \dot{\xi}^a \tilde{x}_a +c(t) \right) \ ,
		\end{equation}
		showing that the transformation causes a shift in the phase, as expected.

		\bigskip
		
		We must now introduce a transformed energy functional so as to allow the entropy to participate in the dynamics, but we cannot assume that either the energy functional or the conservation condition take the same form as in the original frame (\ref{eq:energy1}, \ref{eq:energy_cons}). Rather, we start with the original conservation of energy condition (\ref{eq:energy_cons}) and energy functional (\ref{eq:energy1}). Upon expressing the current velocity in terms of the transformed coordinates ($\smash{v^2 = \tilde{v}^2 - 2\dot{\xi}^a\tilde{v}_a + \dot{\xi}^2}$) and simplifying,
		\begin{eqnarray}
			\label{eq:trans_energy2}
			\dot{E} - \int \! dx\ \rho \dot{V} &=& \int \! dx \ \bigg[ \dot{\rho} \left(\tfrac{1}{2}m \tilde{v}^2 + \tfrac{1}{2}m u^2 + V -m\dot{\xi}^a\tilde{v}_a \right) \nonumber \\
			&&\qquad+\ \rho\left( \tfrac{1}{2}m\partial_t u^2 + mv_a\partial_t \tilde{v}^a \right)\bigg] - \int \! dx\ \rho m\ddot{\xi}^a v_a  = 0 \ .
		\end{eqnarray}
		
		The $\ddot{\xi}$ integral in (\ref{eq:trans_energy2}) arises from the velocity cross term and is of particular interest. Inserting the untransformed current velocity (\ref{eq:untrans_current_velocity}), integrating by parts, and substituting the original FP equation (\ref{eq:fp}) results in
		\begin{eqnarray}
			\label{eq:grav_integral}
			\int \! dx\ (\eta\ddot{\xi}^a) (\rho \partial_a\phi) &=& -\int \! dx \ \eta\ddot{\xi}^a(x_a +d_a(t)) \ \partial^b(\rho \partial_b\phi) \nonumber \\
			&=& \int \! dx \ \dot{\rho} \, m \ddot{\xi}^a\left(\tilde{x}_a + d_a(t)\right) \ ,
		\end{eqnarray}
		where $d^a$ is a constant of integration; its arbitrariness reflects the freedom in choosing the zero of the gravitational potential, which is a kind of gauge transformation. This is not unique to the EQD version of this transformation but exists in the standard formulation as well.
		
		The rest of (\ref{eq:trans_energy2}) is handled in the same way as the original energy condition (\ref{eq:energy1}). Requiring the condition to hold for arbitrary choices of $\tilde{\rho}$ and $\tilde{\phi}$ and manipulating with integration by parts and the FP equations gives
		\begin{equation}
			\label{eq:trans_pde_phi}
			\hbar\tilde{\partial}_t \tilde{\phi} + \frac{\hbar^2}{2m} (\tilde{\partial}_a \tilde{\phi})^2 + \widetilde{V} -\frac{\hbar^2}{2 m} \frac{\tilde{\nabla}^2 \tilde{\rho}^{1/2}}{\tilde{\rho}^{1/2}} = 0 \ ,
		\end{equation}
		where $\hbar = m\sigma^2/\tau$, and the transformed potential energy is
		\begin{equation}
			\label{eq:full_grav_pot}
			\widetilde{V} = V - m \ddot{\xi}^a \left(\tilde{x}_a + d_a(t)\right) \ .
		\end{equation}
		The additional term in the potential is an effective gravitational potential arising from the acceleration of the frame $\smash{\widetilde{\mathcal{X}}}$. The standard interpretation of this potential is that it enters via the strong equivalence principle \cite{rosen:1972}. In EQD, however, the general covariance implied by the strong equivalence principle is the result of an \emph{equivalence of information}. EQD does not require that there be some deeper connection between gravitational potentials and non-inertial frames; it simply states that when making inferences about the dynamics of a quantum particle, gravitational potentials and accelerating frames are informationally equivalent.
		
		By combining this result with the transformed FP equation (\ref{eq:trans_fp}) into a complex function $\smash{\widetilde{\Psi} = \tilde{\rho}^{1/2} e^{i\tilde{\phi}}}$, we obtain the Schr\"odinger equation in terms of the transformed quantities.
		To determine the integration constant $c(t)$ in the entropy shift (\ref{eq:entropy_shift_mostly}), we simply express (\ref{eq:trans_pde_phi}) in the original coordinates. This results in a differential equation with solution
		\begin{equation}
			\label{eq:c_full}
			c(t) = -\frac{1}{2} \int \! dt \ \left(\dot{\xi}^2 - 2\ddot{\xi}^a d_a(t) \right) \ .
		\end{equation}

		Since the choice of gauge is arbitrary and one can always choose a new $\smash{\widetilde{\Psi}}$ differing by a phase so as to eliminate $d^a(t)$, we set it to $0$. This choice is standard and simplifies the form of the phase shift (\ref{eq:c_full}) and the potential (\ref{eq:full_grav_pot}),
		\begin{equation}
			c(t) = -\frac{1}{2} \int \! dt \ \dot{\xi}^2 \ , \qquad \widetilde{V} = V - m \ddot{\xi}^a \tilde{x}_a \ .
		\end{equation}

		Up to a gauge freedom, we have determined the difference between the entropies of the observers,
		\begin{equation}
			\label{eq:entropy_diff_final}
			\widetilde{S}(\tilde{x},\tilde{t}) - S(x,t) = \frac{m}{\hbar} \left( \dot{\xi}^a \tilde{x}_a - \frac{1}{2} \int dt \ \dot{\xi}^2\right) \ .
		\end{equation}
		Again, this difference is also the phase shift between the frames, and it is the same result that the EGT yields in the standard formulation of QM. In entropic dynamics, however, it takes a new meaning as the relation between the states of knowledge which two observers \emph{must} have in order for the extended Galilean transformation to qualify as a symmetry.
				

		The first term in the entropy shift (\ref{eq:entropy_diff_final}) is required for the momentum to transform properly. The second term, however, tells a more interesting story. Dividing by $c^2$ and rearranging, the integral can be written as
		\begin{eqnarray}
			\frac{1}{2c^2} \int \! dt \ \dot{\xi}^2 = t - \int \! dt \ \left( 1 - \frac{\dot{\xi}^2}{2c^2} \right) \approx t - \int \! dt \ \sqrt{ 1 - \frac{\dot{\xi}^2}{c^2} } \ .
		\end{eqnarray}
		The integral in the second step is a first-order approximation of the proper time of the moving observer \cite{greenberger:1979}. We see that the second term of the entropy shift is a residue of special relativity due to the difference in proper time between the two frames of reference. Although our non-relativistic formulation of quantum mechanics makes no distinction between coordinate and proper time, this residual effect indicates that a relativistic quantum entropic theory would necessarily include divergent definitions of time in order to properly reflect the observers' differing states of information.
		
		The transformed dynamical equation (\ref{eq:trans_pde_phi}) has the same form as the original (\ref{eq:pde_phi}). One could obtain this same result starting with a transformed energy functional and conservation condition,
		\begin{equation}
			\label{eq:trans_energy_final}
			\widetilde{E}[\tilde{\rho},\tilde{v}] = \int \! d^3\tilde{x} \ \tilde{\rho}(\tilde{x},\tilde{t}) \left( \tfrac{1}{2}m\tilde{v}^2 + \tfrac{1}{2}m \tilde{u}^2 + \widetilde{V}(\tilde{x},\tilde{t}) \right) \ , \qquad \tilde{\partial}_t \widetilde{E} = \int \! d\tilde{x} \ \tilde{\partial}_t\widetilde{V} \ .
		\end{equation}
		This implies that although the energy and its rate of change in the two frames is very different (as illustrated by the transformed potential (\ref{eq:full_grav_pot})) the energy functional does indeed take the same form, and the conservation requirement is covariant.
		
	\section{Conclusions}
		The concept of symmetries is extremely useful in physics, and their role in entropic dynamics is no exception. For a transformation relating two observers to qualify as a symmetry, the observers must have \emph{equivalent} states of knowledge. This equivalence of information is reflected by the symmetry conditions (\ref{eq:sym_con1}, \ref{eq:sym_con2}) and, ultimately, the invariance of the dynamical equations (\ref{eq:trans_fp}, \ref{eq:trans_pde_phi}). For the EGT, the particular way that the observers' information about the hidden variables must differ manifests itself as an entropy shift (\ref{eq:entropy_diff_final}). Finally, accounting for an effective gravitational potential that exemplifies the equivalence principle leads to in an covariant energy conservation condition (\ref{eq:trans_energy_final}).

	\bibliographystyle{apsrev4-1}
	\bibliography{nrgeqd-maxent-arxiv}

\end{document}